\documentclass[review]{elsarticle}

\usepackage{lineno,hyperref}
\modulolinenumbers[5]
\usepackage{setspace}
\usepackage{epsfig}
\usepackage{delarray}
\usepackage{amsmath, amssymb, amsfonts}
\usepackage{bm}
\usepackage{graphicx}
\usepackage{dcolumn}
\usepackage{float}

\begin{document}
\title{Quantum Fisher Information of an Open and Noisy System in the Steady State}

\author{Azmi Ali Altintas}
 \address{Faculty of Engineering and Architecture, Okan University, Istanbul, Turkey}
\ead{altintas.azmiali@gmail.com}
\date{\today}

\begin{abstract}
We study the  quantum Fisher information (QFI) per particle of an open (particles can enter and leave the system) and  dissipative (far from thermodynamical equilibrium) steady state system of two qubits in a noise which is decoherence. We show the behavior of  QFI per particle of the system due to changes of reset and decoherence parameters r and $\gamma$ respectively. The parameter r is the strength of the reset mechanism,  $\gamma$ is the strength of decoherence and in our case it is dephasing channel. The parameters $\gamma$ and $r$ are real numbers. We observe that the reset parameter must be bigger than decoherence parameter.  We have found that by choosing coupling parameter g as $5\gamma$ the QFI per particle is 1.00226 which is greater than shot noise limit at $\gamma=0.5$ and $r=14$ . Also the concurrence and negativity of the such state have been calculated and they are found as 0.0992486 and 0.0496243 respectively. We have shown that when the concurrence and negativity of some specific states different than zero, which means the state is entangled, the QFI of the system is greater than 1. The QFI per particle, concurrence and negativity shows that the chosen case is weakly entangled.   We discovered that the optimal direction depends on the parameters $r$ and $\gamma$ and a change in the direction effects the behavior of the QFI of the system. 
\end{abstract}
%\Pacs{03.67.Mn, 03.65.Ud, 03.67.-a, 03.65.Ta}

\maketitle

%%%%%%%%%%%%%%%%%%%%%%%%%%%%%%% main %%%%%%%%%%%%%%%%%%%%%%%%%%%%%\begin{figure}[htbp]
%\begin{center}
%\epsfile{file=,scale=0.8}
%\caption{{\bf default}}
%\label{default}
%\end{center}
%\end{figure}
%%%%%%%%%%%%%%%%%%%%%%%%%%%%%%%%%%%%%%%%%%%%%%%%%%%%%%%%%%%%%%%%%%%%%%
\section{Intoduction}
Quantum Fisher information (QFI) which characterizes the sensitivity of a quantum system with respect to the changes of a parameter of the system has been shown to be a multipartite entanglement criteria\cite{PezzeSmerzi2009PRL,Hyllus2012PRA}: If the mean quantum Fisher information per particle of a state exceeds the so called \textit{shot-noise limit} i.e. the ultimate limit that separable states can provide, then the state is multipartite entangled. \textit{Shot-noise limit} is $\Delta \theta \equiv \frac{1}{\sqrt{N}}$, where N is the number of particles\cite{Xiong201010}.  The converse is not generally true that not all pure multipartite entangled states achieve this limit, i.e. they are not \textit{useful for sub-shot-noise interferometry} even if optimized by local operations. The only exception is for N=2 case, at that case any entangled state can be made useful by local operations \cite{Hyllus2010PRA}. It is also shown that GHZ states provide the largest sensitivity, achieving the fundamental, so called Heisenberg limit \cite{Lloyd2004Science}. Mean QFI determines the phase sensitivity of  state with respect to SU(2) rotations. Recently the quantum Fisher information has been further studied both theoretically and experimentally \cite{Ma2011PRA,Ji2008IEEE,Escher2011NatPhys,Wang2012IJTP,Spagnolo2012SciRep,Liu2013IJTP,Ozaydin2013IJTP,Ozaydin2014IJTP,Ozaydin2014ACTA,Ozaydin2014PLA,Gibilisco2007JMP,Andai2008JMP,Berrada2012,Luo2003PRL,Kacprowicz2010NatPhoton,Krischek2011PRL,Ozaydin2015IJTP,Strobel2014Science,Erol2014SciRep,Jing2015PRA,Ozaydin2015ACTA,Ozaydin2015SciRep}.  Quantum Fisher information is mainly related with quantum estimation theory and there are some recent works about quantum estimation in open systems \cite{Gambetta2001PRA,Alipour2014PRL}.

In nature it is hard to  find controlled and closed systems. Usually the natural systems are open and noisy.  If a quantum system interacts with environment it is thought as the quantum system is in a noisy channel. It is known that  the entanglement of the quantum system decreases when the system is in a noisy channel.   However for an open quantum system the decrease in entanglement can be balanced by a reset mechanism. With the help of reset mechanism an entangled steady state can be established. The reset mechanism replaces randomly particles of the quantum system with particles from the environment in some standard, sufficiently pure, single-particle state \cite{Briegel2006PRA}. If the reset mechanism is taken alone, it can not produce entanglement to the system. To create entanglement, the fresh particles must interact with the system. Since there is particle transfer from environment, the system must be open. Hartmann et.al shown that for both gas type and strongly coupled quantum systems the effect of decoherence can be vanished with the help of reset mechanism \cite{Hartmann2007NJP}.  
    
In this work, we study the quantum Fisher information per particle of open and dissipative noisy system of two qubits with reset mechanism. Because of its simplicity dephasing channel is  used as decoherence channel in this study. It  should also be noted that the quantum system is in a steady state.   We examine the effect of reset mechanism if the strength of decoherence $\gamma$ taken to be constant.  We observe that  for the separability of the system, the strength of reset ``r" must be well chosen to balance the effect of decoherence. Then by removing the restriction on $\gamma$ we have looked for the quantum Fisher information of the quantum system and it is observed that the system remains separable by proper combinations of  $\gamma$ and r.  
\section{QFI of open noisy system in a steady state}
The estimating parameters of a quantum state is one of the  tasks of quantum information theory. Let $\phi$ be a parameter of a density matrix $\rho(\phi)$. The quantum Cram\'{e}r-Rao bound is the bound for the variance of estimation of parameter $\phi$.
\begin{equation}
\Delta\phi_{QCB}=\frac{1}{\sqrt{N_mF}},
\end{equation}
where $N_m$ is number of experiments and F is quantum Fisher Information. 
We consider that the parameter $\phi$ is obtained by SU(2) rotations.
\begin{equation}
\rho_\phi=U_\phi\rho U^{\dag}_\phi,
\end{equation}
where $U_\rho=e^{i\phi}J_{\overrightarrow{n}}$ and and $J_{\overrightarrow{n}}=\sum_{\alpha=x,y,z} \frac{1}{2}n_\alpha \sigma_\alpha$, the angular momentum operator in $\overrightarrow{n} $ direction. $\sigma_\alpha $ are Pauli matrices. 

In quantum metrology there are many methods to calculate the quantum Fisher information. One of the methods considered by Liu et.al \cite{Liu2013PRA} is a useful one for separable states. Another approach is about the general parametrization process $U=e^{-itH}$ for an initial pure state with time independent Hamiltonian H depending a parameter $\phi$\cite{Jing2015PRA}. The third method gives the formula for the calculation of quantum Fisher information for density matrices with arbitrary ranks\cite{Liu2014CTP}. 

In this study, we are looking for points where the state is entangled (not separable) by calculating QFI of the system.  The density matrix is a full rank matrix and the Hamiltonian of the system does not depend on the su(2) parameters. In the light of these arguments we use different method than these three methods. The method that we use to calculate the quantum Fisher information can be written from \cite{Hyllus2010PRA} as;
\begin{equation}\label{QFI}
F(\rho,J_{\overrightarrow{n}})=\sum_{i\neq j}\frac{2(p_i-p_j)^2}{p_i+p_j}{|\langle i |J_{\overrightarrow{n}}| j \rangle|}^2=\overrightarrow{n}\textbf{C}\overrightarrow{n}^T.
\end{equation}
Here $p_i$ and $| j \rangle$ are the eigenvalue and eigenvector of state $\rho$ respectively. Also $\overrightarrow{n}$  is a normalized three dimensional vector and  $p_i+p_j=0$ terms are not included to summation.  After some calculations the matrix elements of the symmetric matrix \textbf{C} can be found as;
\begin{equation}
\textbf{$C_{kl}$}=\sum_{i\neq j}\frac{(p_i-p_j)^2}{p_i+p_j}[\langle i | J_{k}|j \rangle\langle j | J_{l}|i \rangle+
\langle i | J_{l}|j \rangle\langle j | J_{k}|i \rangle]
\end{equation}
If $\rho$ is a pure state the equation 4 is written as
\begin{equation}
\textbf{$C_{kl}$}=2\langle J_kJ_l+J_lJ_k\rangle-4\langle J_k\rangle\langle J_l\rangle,
\end{equation}
and the QFI is also expressed as $F(\rho,J_{\overrightarrow{n}})=4(\Delta J_{\overrightarrow{n}} )^2$.
The mean QFI is found as in \cite{Ma2011PRA}
\begin{equation}
\overline{F}_{max}=\frac{F_{max}}{N}=\frac{\lambda_{max}}{N}
\end{equation}
here $\lambda_{max}$ is maximum eigenvalue of matrix C and N is the number of particles. 
It has recently been shown that,  the QFI for separable states is \cite{PezzeSmerzi2009PRL}
\begin{equation}
\overline{F}_{max}\leq 1 
\end{equation} 
and 
for general  states the mean QFI of the system is
\begin{equation}
\overline{F}_{max}\leq N
\end{equation} 
where the bound $\overline{F}_{max}=N$ can be saturated by maximally entangled states.

 Now, we define an open quantum system with reset mechanism in a noisy channel. Since there is some particle transfer the system varies with time. The total master equation which defines the quantum system is given by \cite{Briegel2006PRA}
\begin{equation}\label{master}
\dot{\rho}=-i[H,\rho]+\textit{$L_{noise}\rho$}+r\sum_{i=1}^N(|\chi_i\rangle\langle\chi_i|tr_i\rho-\rho)
\end{equation}
The master equation is in form of Lindbald equation. The solution of the master equation gives us density matrix $\rho$ of the quantum system. In equation (\ref{master}), $|\chi_i\rangle$ is a specific state and we choose it as $|+\rangle_i$ where $\sigma_x^i|+\rangle_i=|+\rangle_i$. 

 The first term in the right hand side of eq. (\ref{master}) is just about total Hamiltonian of the quantum system, the second term describes the noisy channel and the third term describes the reset mechanism and N is the number of particle. Now let us examine  the terms of the right hand side of eq. (\ref{master}) in detail.

   The Hamiltonian of two qubit steady state can be written as,  
 \begin{equation}
 H=g\sigma^{(1)}_z\sigma^{(2)}_z,
 \end{equation}
here $g\geq 0$ is the coupling strength. 

The expression of decoherence channel which is chosen as dephasing channel is
\begin{equation}
\textit{$L_{noise}\rho$}=\frac{\gamma}{2}\sum_{i=1,2}(\sigma_{z}^{(i)}\rho\sigma_{z}^{(i)}-\rho)
\end{equation}
here $\gamma$ is strength of decoherence which is a positive real number and $\sigma_{z}^{(i)}$ is the Pauli spin matrix in z axis for i'th particle.

The  last term in equation (\ref{master}) is about reset mechanism. The expression means that with some probability $r\delta t$ the particle i, $i=1,\cdots, N,$ is resetted during time interval $\delta t $ to some specific state $|\chi_i\rangle$, while the other qubits are stays in the state $tr_i\rho$, i.e. the reduced density matrix obtained by tracing out the i'th particle.  r is the strength of reset mechanism which is a positive real number.  Since the reset state should be able to produce entanglement from the resulting product state,  the reset state must be depend on the Hamiltonian. For example for our Hamiltonian we can not choose the reset state as $|\chi_i\rangle=|0\rangle$, since the state does not create any entanglement. 
Then our two qubit master equation becomes
\begin{equation}\label{master2}
\dot{\rho}=-i[H,\rho]+\frac{\gamma}{2}\sum_{i=1,2}(\sigma_{z}^{(i)}\rho\sigma_{z}^{(i)}-\rho)+r\sum_{i=1,2}(|+\rangle_i\langle +|tr_i\rho-\rho).
\end{equation}
When $r=0$, dephasing part $(\gamma)$ destroys all the entanglement. When $r\rightarrow \infty$ Hamiltonian and noise parts are neglected and since the reset part injects fresh particles to the system the entanglement between two qubits will be zero.
 $\rho$ is the density state and it can be expressed as matrix form. In our case it is $4\;\mbox{by}\;4$ matrix. The matrix is written from \cite{Briegel2006PRA}, its diagonal elements are equal to $\frac{1}{4}$. It means that
\begin{equation}
\rho_{11}=\rho_{22}=\rho_{33}=\rho_{44}=\frac{1}{4}.
\end{equation}
 The anti-diagonal elements are
\begin{equation}
\rho_{14}=\rho_{23}=\rho_{32}=\rho_{41}=\frac{r^2(r+\gamma/2)}{4(r+\gamma)[2g^2+(r+\gamma/2)(r+\gamma)])}.
\end{equation}
The other elements of density state matrix are;
\begin{eqnarray}
\rho_{12}&=&\rho_{13}=\rho_{42}=\rho_{43}=\frac{r(-ig+r+\gamma/2)}{4[2g^2+(r+\gamma/2)(r+\gamma)]},\\
\rho_{21}&=&\rho_{24}=\rho_{31}=\rho_{34}=\frac{r(ig+r+\gamma/2)}{4[2g^2+(r+\gamma/2)(r+\gamma)]}.
\end{eqnarray}
Here r, $\gamma$ and g are reset, decoherence and coupling strength parameters respectively and they are real parameters.

Now by using $\rho$ matrix in equation (\ref{QFI}) we find  QFI per particle of the system depending on parameters r, $\gamma$ and g. 

 If the coupling parameter is chosen as smaller than noise parameter, the decoherence would destroy the entanglement. It means that we should take the coupling parameter is bigger than noise parameter $\gamma$. 
 
 In our study we take  $g=5\gamma$ then the QFI depends only on reset and decoherence parameters.  One can construct the QFI matrix C and it is in the form of
 \begin{equation}C=
 \left(\begin{array}{ccc}
 C_{xx} & 0 & 0 \\
 0 & C_{yy} & C_{yz} \\
 0 & C_{zy} & C_{zz}
 \end{array}
 \right),
 \end{equation} where $C_{yy}=C_{zz}\;\mbox{and } C_{yz}=C_{zy}$. The form of matrix means that the rotation is in x direction. 

  To understand the behavior of the  QFI per particle first we have fixed the noise parameter to $\gamma=0.5.$

When the reset parameter r equals to zero the nonzero elements of the density matrix $\rho$ are only in diagonal  as $1/4$.  Then the quantum Fisher information of this density matrix is zero. The result is expected since when $r=0$ the noise destroys entanglement. Also when the reset parameter goes to infinity, the non zero elements of $\rho$ are in diagonal and anti-diagonals as $1/4$ so one calculate that the QFI of the density matrix is also zero. The result is reasonable, since the reset mechanism replaces particles with fresh particles, when $r \rightarrow \infty$ the system is not able to establish entanglement so the QFI will be zero.

\begin{figure}[H]
\includegraphics[width=1\textwidth]{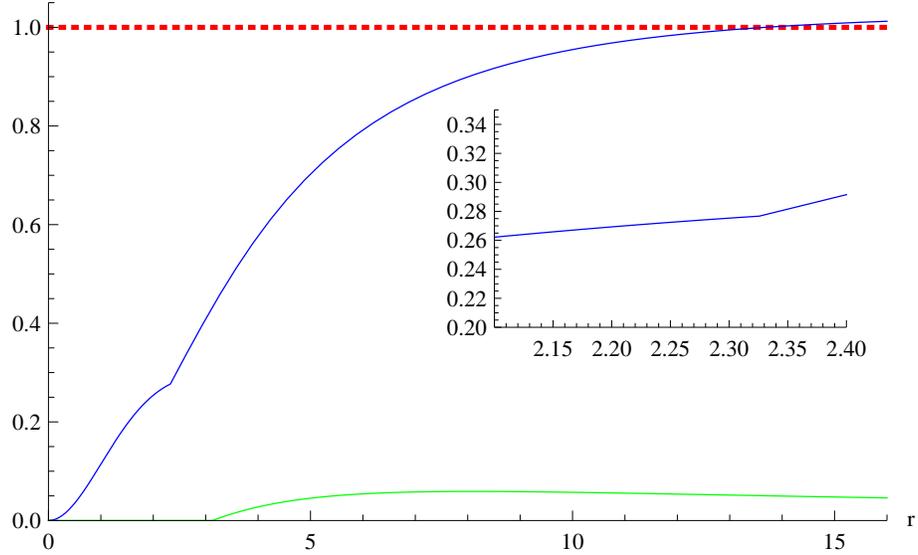}
\caption{ QFI per particle (blue) and negativity (green) vs reset with $\gamma=0.5$. Red dotted line represents the shot noise limit. Inset shows the critical point where the optimal direction changes. } \label{fig:Fisher}
\end{figure}
As one can see from the figure at  $r=0$ the QFI of the system is 0 as expected. When r is at 14  QFI per particle has a value as 1.00226. Well known entanglement criteria are concurrence and negativity and both can take values between 0 and 1. For our entangled state  concurrence is 0.0992486 and negativity is 0.0496243. It means that the chosen state is weakly entangled. 
When negativity is 0 the state is separable, when negativity equals to 1 the state is maximally entangled. Same conditions are hold for concurrence. 

A critical point is observed at $r=2.3$, at that point the behavior of QFI is changing. The reason of this change could be change in the optimal direction.  It is well known that the optimal direction $\overrightarrow{n}^o$ is determined by the eigenvector of symmetric matrix C with the maximal eigenvalue.  We discovered that the optimal direction changes with the values of $\gamma$ and r. For the figure (1) ($g=2.5$ and $\gamma=0.5$ case) the optimal direction $\overrightarrow{n}^o=\overrightarrow{n}_1$ when $r \leq 2.3$. For $r>2.3$ the optimal direction  $\overrightarrow{n}^o=\overrightarrow{n}_2\sin(\pi/4)+\overrightarrow{n}_3\cos(\pi/4)$. Here $\overrightarrow{n}_1$ is unit vector in x axis. $\overrightarrow{n}_2\; \mbox{and}\; \overrightarrow{n}_3$ are unit vectors in y and z directions respectively.

Then to understand the system better, we cancel the restriction on $\gamma$, and we have examined the case $\gamma$ takes values between 0.01 and 3.  If the decoherence parameter started from 0 the QFI per particle would be 0 and we could not observe any entangled case. Since we want to observe the decay of QFI per particle $\gamma$ starts from 0.01.

 \begin{figure}[H]
\includegraphics[width=1\textwidth]{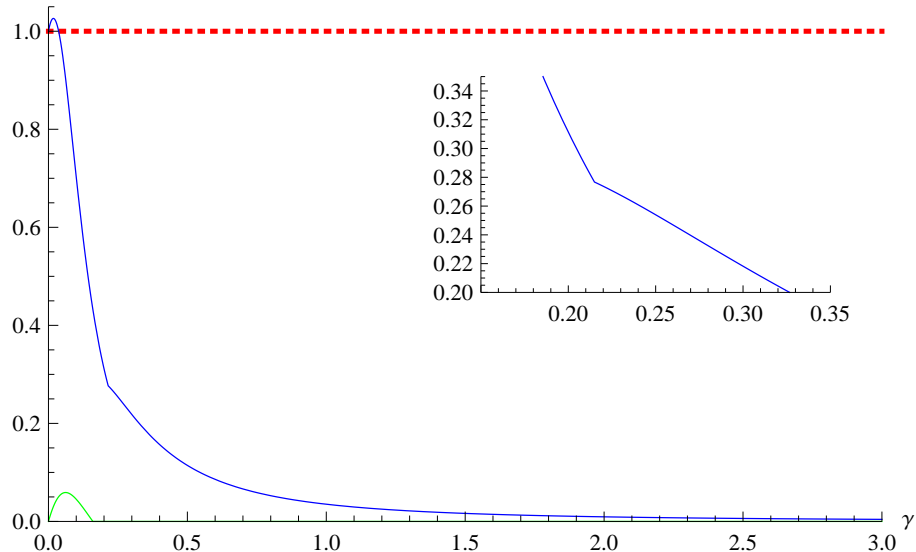}
\caption{QFI per particle (blue) and negativity (green) vs reset with $r=1$. Red dotted line represents the shot noise limit. Inset shows the critical point where the optimal direction changes.  } \label{fig:Fisher1}
\end{figure}
 In figure (2) the QFI per particle is 1.02124 when $\gamma=0.01$. While the negativity of this specific state is 0.0183813, the concurrence is 0.0367627.  It means that with the  proper combinations of reset, decoherence and coupling parameters one can find entangled states. Also  it can be seen that there is a critical point at $\gamma=0.214$. This is related with optimal direction, because at that point the optimal direction is $\overrightarrow{n}^o=\overrightarrow{n}_2\sin(\pi/4)+\overrightarrow{n}_3\cos(\pi/4)$ and above that point the optimal direction is $\overrightarrow{n}^o=\overrightarrow{n}_1$. 
 
Both figures show that when negativity is different than zero  the QFI is not greater than 1 simultaneously. It means that the QFI does not recognize all the entangled states.  

 Figure (1) and (2) give two important results. The first one, by choosing parameters g, $\gamma$ and r properly one can get entangled states. The second one, the optimal direction depends on the parameters of the system and the change in the optimal direction changes the behavior of the QFI per particle of the system.
 
% \begin{figure}[H]
% \includegraphics[width=1.0\textwidth]{fig3d2.eps}
% \caption{(Change in QFI per particle of the system with respect to r and $\gamma$.) }  % \label{fig:2}
% \end{figure}
\section{Conclusion}

 We know that entanglement of any quantum system decreases in a noisy channel. To prevent the decrease in entanglement a reset mechanism can be used\cite{Briegel2006PRA}. Since the reset mechanism changes the particles of the system with fresh particles from the environment the quantum system must be open and obviously the system is far from thermodynamical  equilibrium. 
  
 We have studied the quantum Fisher information of a noisy open quantum system of two qubits which is in a steady state. In the system, we use an interaction Hamiltonian in a dephasing channel and a reset mechanism. By solving master equation we have defined density state in matrix form and with the help of equation (\ref{QFI}) we describe mean QFI of the system depending on reset and noise parameters. In this work the coupling strength has been chosen as $g=5\gamma$ thus the QFI only depends on reset and noise parameters. Since mean QFI characterizes the phase sensitivity of a state with respect to SU(2) rotation, by finding the general shape of QFI matrix and we have found that the rotation is around x axis. 
 
 Then we have shown the change of  QFI per particle depending on these parameters in two figures.  If the system is in a constant noise (see figure 1) when the power of reset mechanism is increased the usefulness of the state is also increasing. In figure 1 we have chosen $\gamma$ as $0.5$ and have observed that at $r=14$ QFI per particle is greater than 1. The concurrence and negativity of such state is greater than 0. It means that the state is entangled. However the figure is also shows that there are some states although the negativity and concurrence of them are different than zero, the QFI per particle is less than 1.

  The second figure shows the behavior of  QFI of the system when reset parameter is equal to 1 and noise parameter is  changing between 0 and 3. We can find entangled states by choosing $\gamma $ properly. For example, when $\gamma=0.01$ the QFI per particle is 1.02124. For this case the concurrence and negativity is different from zero. 
 
At some specific states concurrence, negativity and QFI per particles show that the state is entangled. Explicitly figure (1) shows that there are some states when negativity is different than zero the QFI of the state is below the shot-noise limit. It means that QFI per particle does not recognize all the entangled states.   
 
 The optimal direction $\overrightarrow{n}^o$ depends on the reset and noise parameters. The optimal direction found from the constant noise is opposite to the optimal direction for the constant reset case. This is an important finding which shows the change in optimal direction affects the behavior of the QFI of the system.  
  
 As a further study one can look at the quantum Fisher information of an open dissipative system in  a noisy channel which is different than dephasing channel. Also, more than 2 qubit case is an interesting problem. The results of these two problems may give an idea about how to maintain the entanglement in some noisy channels.  
\section*{Acknowledgment}
This work was funded by Isik University Scientific Research Fund, Grant Number: BAP-14A101


\begin{thebibliography}{99}

\bibitem{PezzeSmerzi2009PRL}
 Pezze L., Smerzi A., Entanglement, Non-linear Dynamics, and the Heisenberg Limit. Phys. Rev. Lett. \textbf{102,} 100401 (2009).


\bibitem{Hyllus2012PRA}
Hyllus, P.,  et al.,Fisher Information and Multiparticle Entanglement. Phys. Rev. A \textbf{85,} 022321 (2012).

\bibitem{Xiong201010}
Xiong, H. N., Ma, J. , Liu, W. F.  and Wang,  X., Quantum Fisher Information for Superpositions of Spin States.  Quant. Inf. Comp. \textbf{10,} 5\&6 (2010).

\bibitem{Hyllus2010PRA}
Hyllus, P., Gühne,  O. and Smerzi, A., Not All Pure Entangled States are Useful for Sub Shot-Noise Interferometry. Phys. Rev. A \textbf{82,} 012337 (2010).

\bibitem{Lloyd2004Science}
Giovannetti, V.,  Lloyd, S., and Maccone, L., Quantum-Enhanced Measurements:Beating the Standard Quantum Limit. Science, \textbf{306,} 1330 (2004).

%QFI
\bibitem{Ma2011PRA}
Ma, J., Huang, Y., Wang, X. and Sun, C. P. , Quantum Fisher Information of the Greenberger-Horne-Zeilinger State in Decoherence Channels. Phys. Rev. A, \textbf{84,} 022302 (2011).


\bibitem{Ji2008IEEE}
Ji, Z.  et. al, Parameter Estimation of Quantum Channels. IEEE Trans. Info. Theory, \textbf{ 54,} 5172 (2008).

\bibitem{Escher2011NatPhys}
Escher, B. M., Filho, M. and Davidovich, L., General Framework for Estimating the Ultimate Precision Limit in Noisy Quantum-Enhanced
Metrology.  Nat. Phys., \textbf{7,} 406 (2011).

\bibitem{Wang2012IJTP}
Yi, X., Huang, G.  and Wang, J., Quantum Fisher Information of a 3-Qubit State. Int. J. Theor. Phys. \textbf{51,} 3458 (2012).

\bibitem{Spagnolo2012SciRep}
Spagnalo, N.  et al.,Quantum Interferometry With Three-Dimensional Geometry. Sci. Rep.,\textbf{ 2,} 862 (2010).

\bibitem{Liu2013IJTP}
Liu, Z., Spin Squeezing in Superposition of Four-Qubit Symmetric State and W States. Int. J. Theor. Phys., \textbf{52,} 820 (2013).

\bibitem{Ozaydin2013IJTP}
Ozaydin, F. , Altintas, A. A., Bugu, S. and Yesilyurt, C., Quantum Fisher Information of N Particles in the Superposition of W and GHZ States. Int. J. Theor. Phys.,  \textbf{52,} 2977 (2013).

\bibitem{Ozaydin2014IJTP} Ozaydin, F. , Altintas, A. A., Bugu, S. and Yesilyurt, C., Quantum Fisher Information of N Particles in the Superposition of W and GHZ States. Int. J. Theor. Phys., \textbf{53,} 3219 (2014).

\bibitem{Ozaydin2014ACTA}
Ozaydin F. , Altintas A. A., Bugu S. and Yesilyurt C., Behavior of Quantum Fisher Information of Bell Pairs Under Decoherence Channels. Acta Physica Polonica A, \textbf{125,} 606 (2014).

\bibitem{Ozaydin2014PLA}
Ozaydin F., Phase Damping Destroys Quantum Fisher Information of W states Phys. Lett. A, \textbf{378}, 43 (2014).

\bibitem{Gibilisco2007JMP}
Gibilisco, P. , Imparato, D., and Isola, T.,  Uncertainty Principle and Quantum Fisher Information. II. J. Math. Phys., \textbf{48,} 072109 (2007).

\bibitem{Andai2008JMP}
Andai, A., Uncertainty Principle with Quantum Fisher Information. J. Math. Phys.,  \textbf{49,} 012106 (2008).
\bibitem{Berrada2012}
Berrada, K., Khalek, S. B.  and Obada, A.S.F., Quantum Fisher Information for a Qubit System Placed Inside a Dissipative Cavity.  Phys. Lett. A \textbf{376}, 1412 (2012).

\bibitem{Luo2003PRL}
Luo, S., Wigner-Yanase Skew Information and Uncertainty Relations. Phys. Rev. Lett., \textbf{91,} 180403 (2003). 

\bibitem{Kacprowicz2010NatPhoton}
Kacprowicz, M.  et al., Experimental Quantum-Enhanced Estimation of a Lossy Phase Shift. Nat. Photon. \textbf{4,} 357 (2010).

\bibitem{Krischek2011PRL}
Krischek, R. et al., Useful Multiparticle Entanglement and Sub-Shot-Noise Senitivity in Experimental Phase Estimation. Phys. Rev. Lett., \textbf{107,} 080504 (2011).

\bibitem{Ozaydin2015IJTP}
Ozaydin. F., Quantum Fisher Information of a 3x3 Bound Entangled State and its Relation with Geometric Discord. Int. J. Theor. Phys., \textbf{54,} 3304 (2015).

\bibitem{Strobel2014Science}
Strobel, H. et al., Fisher Information and Entanglement of non-Gaussian Spin States. Science, \textbf{345,} 424 (2014).

\bibitem{Erol2014SciRep}
Erol V., Ozaydin F. and Altintas, A. A., Analysis of Entanglement Measures and LOCC Maximized Quantum Fisher Information of General Two Qubit Systems, Scientific Reports, \textbf{4,} 5422 (2014).

\bibitem{Jing2015PRA}
Jing X. X., Liu J. Xiong H. and Wang X., Maximal Quantum Fisher Information for General SU(2) Parametrization Process, Phys. Rev. A, \textbf{92,} 012312 (2015).

\bibitem{Ozaydin2015ACTA}
Ozaydin, F. , Altintas A. A., Yesilyurt, C., Bugu S., and Erol V., Quantum Fisher Information of Bipartitions of W States. Acta Physica Polonica A, \textbf{127,} 1233 (2015).

\bibitem{Ozaydin2015SciRep}
Ozaydin, F. and  Altintas A. A., Quantum Metrology: Surpassing the Shot-Nise Limit with Dzyaloshinskii-Moriya Interaction. Scientific Reports, \textbf{5}, 16360, (2015). 

\bibitem{Gambetta2001PRA}
Gambetta, J., Wiseman, H. M., State and Dynamical Parameter Estimation for Open Quantum Systems. Phys. Rev. A \textbf{64,} 042105 (2001). 

\bibitem{Alipour2014PRL}
Alipour, S. Mehboudi, M. and Rezakhani, A. T., Quantum Metrology in Open Systems: Dissipative Cramér-Rao Bound. Phys. Rev. Lett. \textbf{112,} 120405 (2014).

\bibitem{Briegel2006PRA}
 Hartmann, L., D\"{u}r, W., and Briegel, H.J., Steady-State Entanglement in Open and Noisy Quantum Systems. Phys. Rev. A \textbf{74,} 052304 (2006).

\bibitem{Hartmann2007NJP}
  Hartmann, L., D\"{u}r, W., and Briegel, H.J., Entanglement an its Dynamics in Open Dissipative Systems. NJP \textbf{9,} 230 (2007).

\bibitem{Liu2013PRA}
Liu, J., Jing, X., and 	Wang, X., Phase-matching Condition for Enhancement of Phase Sensitivity in Quantum Metrology.  Phys. Rev. A \textbf{88,} 042316 (2013).


\bibitem{Liu2014CTP}
Liu, J., Jing, X., Zhong, W., Wang, X., Quantum Fisher Information for Density Matrices with Arbitrary Ranks. Commun. Theor. Phys. \textbf{61,} 45 (2014).


\end{thebibliography}
\end{document}